\documentclass[conference]{IEEEtran}
\usepackage[utf8x]{inputenc}
\usepackage[T1]{fontenc}
\usepackage[english]{babel}
\usepackage{color}
\usepackage{amsthm}
\usepackage{enumerate}
\usepackage{verbatim}
\usepackage{tikz}
\usetikzlibrary{through,intersections}
\usepackage{fixme} %
\usepackage{cite}
\usepackage{mathpazo}
\usepackage{url}

\usepackage{verbatim, amsmath, amsfonts, amssymb, amsthm}
\usepackage{tikz}
\usetikzlibrary{through,intersections}

\usepackage[kerning=true]{microtype} 
\usepackage[framemethod=TikZ]{mdframed}
\usepackage{bold-extra}

\newtheorem{theorem}{Theorem}
\newtheorem{lemma}{Lemma}
\newtheorem{corollary}{Corollary}
\newtheorem{proposition}{Proposition}

\newtheorem{question}{Question}
\newtheorem{remark}{Remark}
\newtheorem{definition}{Definition}
\usepackage{array}
\usepackage{amssymb}
\usepackage{amsmath}

\newcommand*{\lon}{
       \mskip1mu
        \relax
        {:}
        \mskip1mu
        \relax
}

\newcommand*{\defeq}{\stackrel{\text{def}}{=}}

\title{Dual Information Inequalities}
\title{The Entropy Region is not Closed Under Duality}

\author{
  \IEEEauthorblockN{Tarik~Kaced}
  \IEEEauthorblockA{email: tarik.kaced@ens-lyon.org} 
}

\colorlet{problem}{green!50!yellow}
\colorlet{todo}{red}
\colorlet{toreview}{blue}
\colorlet{standard}{gray}
\colorlet{remark}{green!60!blue}
\colorlet{tokeep}{black!10!violet}
\colorlet{ok}{black}

\DeclareMathOperator{\cl}{\textrm{cl}}
\newcommand{\prob}{\mathrm{Prob}}

\newcommand{\RR}{\mathbb{R}}
\newcommand{\Hent}{\mathbf{H}_N^\text{ent}}
\newcommand{\Hpoly}{\mathbf{H}_N}
\newcommand{\Hrank}{\mathbf{H}_N^\text{Ingl}}

\newcommand{\GF}[1]{\mathbb{F}_{#1}}

\begin{document}
\maketitle
\thispagestyle{plain}
\pagestyle{plain}

\begin{abstract}


  We import a duality notion coming from polymatroids to define duality for
information inequalities. We show that the entropy region for $n\ge 5$ is not
closed under duality.  Our result answers an open question of Matùš
\cite{matus-asc} (1992).

\end{abstract}

\section{Introduction}

Let $(X_i)_{i\in N}$ be $n$ discrete random variables.  To each non-empty
subset of variables $X_J = \{ X_i: i\in J\subseteq N\}$, we can associate the
Shannon entropy $H(X_J)$.  The \emph{entropic vector}
$(H(X_J))_{\varnothing\neq J\subseteq N}$ is a point in the
$2^n-1$-dimensional Euclidean space $\mathbb{R}^{2^n-1}$.  We denote by $\Hent$
the set of all entropic points.  $\Hent$ is a solid object, but not closed in
general \cite{ZY97}.  The closure of $\Hent$ is what we call the entropy
region, it is the set of $2^n-1$-dimensional real vectors that are limits of
entropic vectors.  In fact, $\cl\left(\Hent\right)$ is a convex cone whose
boundary can be delimited by hyperplanes. Such a hyperplane defines a linear inequality
of entropy terms: an information inequality.  Information inequalities live in
the dual\footnote{This duality notion is not the polymatroid duality notion we
are looking for.} cone $(\Hent)^* = \{c\in\RR^{2^n-1} : \langle c,h\rangle
\ge0, h\in \Hent\}$.  A point $c=(c_J)_{\varnothing\neq J\subseteq N}$ in the
dual cone $(\Hent)^*$ commonly corresponds to the coefficients of a linear
information inequality, often rewritten as \[ \sum_J c_J H(X_J)\ge0.  \]
Characterizing the entropy region using information inequalities is no easy
task, a full description is already lacking when $n\ge4$ for it is not
polyhedral \cite{Matus-Inf}.


Fujishige noticed that entropic vectors are in fact polymatroids
\cite{fujishige} in the following sense.
A polymatroid is a real-valued function $f$ defined on
subsets of the ground set $N$ that is non-negative, non-decreasing, and
submodular:
\[
\forall A,B\subseteq N, 
  f(A) + f(B) \ge f(A\cup B) + f(A \cap B).
\]
Indeed, submodularity is related to Shannon's basic inequality:
\begin{equation*}
H(A,C) + H(B,C) \ge H(A,B,C) + H(C),
\end{equation*}
It states that the conditional mutual information $I(A\lon B| C)$ is
non-negative.  The set of positive linear combinations of instances of the
basic inequality are called the \emph{Shannon-type} inequalities.

The duality notion we allude to in the title of this paper comes from
polymatroid theory.  Common operations on the set of polymatroids include
direct sums, linear combinations, minors: deletion and contraction,
convolutions, and duals.  Apart from the last one, all other operations have
entropic counterparts.  We thus concentrate on the under-examined
notion of polymatroid duality.  Let $f$ be a polymatroid on $N$, define the
function \[f^\perp(J) = f(N\setminus J) - f(N) + \sum_{j\in J} f(j).\] Then
$f^\perp$ is again a polymatroid on $N$ and is called the dual polymatroid.  This
operation immediately begs for a similar question for almost entropic vectors: 
the elements of $\cl\left(\Hent\right)$.
\begin{question}
  \label{question-dual}
  Is the dual of an almost entropic polymatroid still almost entropic?
\end{question}

This question is related to open problems about polymatroid duality that can be
found in \cite{matus-asc}.  To attack this question, we propose to shift the
point-of-view to information inequalities.  The key idea is to import the
notion of polymatroid duality into information  inequalities and reformulate
Question~\ref{question-dual} in this setting.  Our main theorem provides a
negative answer to Question~\ref{question-dual} as a corollary.

\begin{theorem}[Main Theorem]
\label{thm:main}
  The entropy region is not closed under duality for $n\ge 5$
\end{theorem}

In the rest of this paper, we provide some properties 
of our duality notion and its connection with balanced inequalities.
We study its meaning for different kinds of information inequalities.

\section{Preliminaries and Properties}

\newcommand{\Ingl}{\mathrm{Ingl}}

For the sake of conciseness we make the following use of notations.
We usually omit commas in entropic terms and by $H(AB)$ we mean $H(A,B)$.  An
instance of an inequality ${\cal I}$ is simply a version of ${\cal I}$ for some
variables assignment.  A conditional version of an inequality ${\cal I}$ is a
version of ${\cal I}$ wherein each entropy term $H(X_J)$ has been replaced by
$H(X_J|Z)$, where $Z$ is a fresh random variable.  If ${\cal I}$ is valid, then
so is its conditional version.  We denote the conditional Ingleton inequality
quantity in the following way:
\begin{multline*}
  \Ingl(A\lon B,C\lon D|E) =\\
I(A\lon B|CE) + I(A\lon B|DE) + I(C\lon D|E) - I(A\lon B|E).
\end{multline*}
The famous
Ingleton inequality \cite{ingleton} thus rewrites as $\Ingl(A\lon B,C\lon D|\varnothing) \ge 0$.

\subsection{Duality and balancing}

\begin{definition}[Balanced Inequalities]
  An $n$-variable information inequality $c\in (\Hent)^*$ is \emph{balanced}
  if the sum of the coefficients involving $X_i$ is zero, for each $i \in N$
  \begin{equation*}
    \forall i\in N,\ \sum\limits_{i\in J\subseteq \mathcal{N}}{c_J} = 0.
\end{equation*}
\end{definition}

Given a valid linear information inequality, 
its balanced counterpart is also valid \cite{chan-balanced}.

\begin{proposition}[Balanced Inequalities, Chan
  \cite{chan-balanced}]\label{theo:balanced}
  Let $(c_J)_{J\subseteq \mathcal{N}}$ be a list of
  coefficients, the following are equivalent:
  \begin{itemize}
    \item The inequality
      \begin{equation}
        \label{eq:bal-unbal}
        \sum_{J\subseteq\mathcal{N}} {c_J H(X_J)}  \ge 0
      \end{equation}
      is a valid information inequality.
    \item The inequality
      \begin{equation}
        \label{eq:bal-bal}
        \sum_{J\subseteq\mathcal{N}  } {c_J H(X_J)} -
        \sum_{ \jmath \in\mathcal{N}  } {r_\jmath H(X_\jmath|
        X_{\mathcal{N}-\jmath}}) \ge 0,
      \end{equation}
      where $r_\jmath$ is the sum of all $c_J$ involving $X_\jmath$,
      is a valid balanced information inequality.
  \end{itemize}
  We say that~(\ref{eq:bal-bal}) is the balanced version of~(\ref{eq:bal-unbal}).
\end{proposition}

We introduce a dual operator for information quantities.

\begin{definition}[Dual operator]
Let $n$ be a number of variables, the dual operator $^\perp$ is defined as
an operator that maps any entropic quantity to a dual quantity by replacing
entropy terms as follows:
\begin{IEEEeqnarray}{r}
\label{eq:dual}
 H^\perp(X_J) \defeq -H(X_J|X_{N\setminus J}) + \sum_{j\in J}H(X_j).
 \end{IEEEeqnarray}
\end{definition}

We are now able to define the formal dual of an information inequality by
defining its dual coefficients.

\begin{definition}[Dual coefficients]
  Let $c=(c_J)_{\varnothing\neq J\subseteq N}$ be the coefficients of an inequality.
  We define the formal dual inequality  $c^\perp$ as the coefficients of the dual of $c$:  
  \[ \left[\sum_J c_JH(X_J) \right]^\perp =
   \sum_J c_JH^\perp(X_J) \defeq 
   \sum_J c^\perp_JH(X_J) \]
\end{definition}

\begin{definition}[Self-dual inequality]
We say that an inequality $c$ is \emph{self-dual} if $c^\perp$ is an
instance of $c$ or a conditional version of $c$.
\end{definition}

We are now able to prove some properties of this duality notion. Let us show it behaves 
as a dual, modulo balancing\footnote{Properties ``modulo balancing'' were first spotted in \cite{equivalence}.}. 

\begin{proposition}
  \label{prop:dualbalanced}
  Let $c\in(\Hent)^*$ be an information inequality, then:
  \begin{enumerate}
    \item
      $c^\perp$ is balanced,
    \item  if c is balanced, then $c^{\perp\perp} = c$,
    \item  if $c$ is not balanced, then $c^{\perp\perp}$ is the balanced version of $c$.
  \end{enumerate}
\end{proposition}
\begin{IEEEproof}
Let $c\in(\Hent)^*$ be an inequality and $c^\perp$ its formal dual.
Let $i\in N$ be a variable index and $J\subseteq N$.

1. We compute the contribution of each term $c_JH^\perp(X_J)$ to the sum $r_i$
of coefficients involving $X_i$. We have
\[c_JH^\perp(X_J) = c_J[H(X_{N\setminus J}) - H(X_N) + \sum_{i\in J}H(X_i)],\]
 if $i\in J$, then the contribution of $c_JH^\perp(X_J)$ is
\[ c_J[0 - 1 + 1] = 0,\]
 if $i\notin J$, then the contribution of $c_JH^\perp(X_J)$ is
\[ c_J[1 - 1 + 0] = 0.\]
Overall, for any $i, r_i = 0$, which implies $c$ is balanced.

2. We compute the contributions of each term $c_JH(X_J)$  
\begin{IEEEeqnarray*}{rCl}
  [c_JH(X_J)]^{\perp\perp} &=& c_J[H^\perp(X_{N\setminus J}) - H^\perp(X_N) + \sum_{i\in J}H^\perp(X_i) ]\\
 H^\perp(X_{N\setminus J}) &=& H(X_J) - H(X_N) + \sum_{i\in J}H(X_i) \\
 H^\perp(X_N) &=& -H(X_N) + \sum_{i\in N}H(X_i)\\
 H^\perp(X_i) &=& -H(X_i|X_{N\setminus\{i\}}) + H(X_i)
\end{IEEEeqnarray*}
By collecting every term we get after cancelling:
      \[
        \sum_{J\subseteq\mathcal{N}  } {c_J H(X_J)} -
        \sum_{i \in\mathcal{N}  } {r_\imath H(X_i|X_{\mathcal{N}-i}}), 
      \]
where $r_\imath$ is the sum of all $c_J$ involving $X_j$. That is we get the balanced version of the original inequality $c$. 

3. The first two properties imply the third one.
\end{IEEEproof}

\begin{remark}
The duality notion from \cite{polyquantoids} is an involution on the set of
polymatroids and induces a involution on all information inequalities which
coincides with ours on balanced information inequalities.
Although it would get rid of the need of balanced inequalities,
their notion is made slightly more complex by additional terms to 
artificially ensure that the ''private information'' of a polymatroid is not
lost.
\end{remark}

\subsection{Polymatroid and vector spaces}

The polymatroid region $\Hpoly$ is the closure of the set of all polymatroids
on the ground set $N$ of size $n$. It is the polyhedral cone delimited by Shannon-type
inequalities.  It is an outer bound of the entropy region.
The following well-known proposition implies that the dual of a polymatroid is
again a polymatroid. Our proof highlights how duality works on basic
inequalities.

\begin{proposition}
  \label{prop:dualPoly}
  The polymatroid region is closed under duality for any $n\ge 0$
\end{proposition}
\begin{IEEEproof}
It suffices to show that the dual of Shannon's basic inequality is valid.
\begin{IEEEeqnarray*}{rCl}
  [ I(A\lon B|C)]^\perp &=& H^\perp(AC) + H^\perp(BC) - H^\perp(ABC) - H^\perp(C) \\ 
                    &=& -H(AC|BD) + H(A) + H(C) -\\
                    &\phantom{= }&-H(BC|AD) + H(B) + H(C) + \\
                    &\phantom{= }& +H(ABC|D) - H(A) - H(B) - H(C) +\\
                    &\phantom{= }& +H(C|ABD) - H(C)\\
                    &=&H(BD) + H(AD) - H(C) - H(ABD)\\
  \left[ I(A\lon B|C)\right]^\perp &=&I(A\lon B|D) \ge 0
\end{IEEEeqnarray*}
We have just proved that Shannon's basic inequality, $I(A\lon B|C) \ge 0$, is self-dual, which implies the result.
\end{IEEEproof}

An interesting subset of entropic vectors is the one arising from vector
subspaces. Let $V$ be a vector space over $\GF{q}$ and
$V_1,V_2,\ldots,V_n\subseteq V$ be $n$ vector subspaces. Denote by $V_J$ the
sum vector subspace $\langle\{V_i,i\in J\}\rangle$, then the point $(\log q
\cdot \dim(V_J))_J$ is an entropic vector \cite{HRSV}. The closure of the set
of all such points is called the Ingleton region $\Hrank$.

\begin{proposition}
  \label{prop:dualIngl}
  The Ingleton region is closed under duality for $n\ge 0$
\end{proposition}

This result is a corollary of the construction of a representation of the dual
polymatroid. A proof of the dual representation can be found in
\cite{padro-ceunotes}, it is based on a matroidal version from Oxley's Matroid
Theory book for matroids \cite{oxley-book}.  Such a construction based on vector
space orthogonality has been used in several constructions related to
information theory (see \cite{fehr-dual,fhkp}).  In our case,
Proposition~\ref{prop:dualIngl} follows from the fact that the dual of an
entropic point arising from vector subspaces is also an entropic point coming
from vector subspaces and Proposition~\ref{prop:dualbalanced}.

Duality induces symmetries that were somehow missed.
For instance, notice that Ingleton inequality is self-dual:
\begin{lemma}
  On variables $A,B,C,D,E,F$, we have:
  \[Ingl^\perp(A\lon B,C\lon D|E) = Ingl(C\lon D,A\lon B|F).\]
  \end{lemma}
  \begin{IEEEproof}
    The theorem statement is the most general but 
    for simplicity we prove the unconditional version on $4$ variables:
 $Ingl^\perp(A\lon B,C\lon D) = Ingl(C\lon D,A\lon B)$. The more general results follows
 from the same type of computations.
\begin{IEEEeqnarray*}{l}
  \phantom{=}  [I(A\lon B|C) + I(A\lon B|D) + I(C\lon D) - I(A\lon B) ]^\perp\\
  = I(A\lon B|D) + I(A\lon B|C) + I(C\lon D|AB) - I(A\lon B|CD)\\
  = I(C\lon D|A) + I(C\lon D|B) + I(A\lon B) - I(C\lon D) 
\end{IEEEeqnarray*}
In the first equation, the dual operator applies linearly, therefore
Proposition~\ref{prop:dualPoly} can be applied.
The last equation is gotten by rearranging the entropy terms.
\end{IEEEproof}

The inequalities on the ranks of five vector subspaces have been studied by
Dougherty~\emph{et al} \cite{dougherty-ranks}. They found $24$ new rank
inequalities on five variables. If we account for duality, the list of $24$
inequalities reduces to $13$, as some inequalities are dual of one another or
self-dual.

\section{Main result}

\begin{lemma}[MMRV inequality \cite{MMRV}]
  The following is a non-Shannon-type information inequality.
  \begin{multline}
    \label{ineq:mmrv}
    I(A\lon B) \le I(A\lon B|C) +I(A\lon B|D)+I(C\lon D)+\\+
    I(A\lon B|E)+I(A\lon E|B)+I(B\lon E|A)
  \end{multline}
\end{lemma}

We are now ready to prove the main theorem.

\begin{IEEEproof}[Proof of Theorem~\ref{thm:main}]
We prove that the formal dual of the MMRV inequality
on five variables is not a valid information inequality.
We provide 
a counter-example by exhibiting a binary joint distribution for variables $A,B,C,D,E$.

Let us first compute the formal dual of inequality~\eqref{ineq:mmrv}.
We first make appear the Ingleton quantity.
\begin{multline*}
    [\Ingl(A\lon B,C\lon D)+I(A\lon B|E)+I(A\lon E|B)+I(B\lon E|A)]^\perp
\end{multline*}
The dual operator acts linearly, so we take the dual of each term and obtain the
following quantity:
\begin{multline*}
\Ingl(C\lon D,A\lon B|E) + I(A\lon B|CD)+I(A\lon E|CD)+I(B\lon E|CD) 
\end{multline*}
which rewrites after expanding the Ingleton term as:
\begin{multline*}
  I(C\lon D|AE) +I(C\lon D|BE)+I(A\lon B|E)- I(C\lon D|E)+\\+
  I(A\lon B|CD)+I(A\lon E|CD)+I(B\lon E|CD).
\end{multline*}

We show that the previous quantity can be negative and thus cannot induce a
valid information inequality.
Consider the distribution on $A,B,C,D,E$ induced by the following tuples with positive
probability masses as shown.
$$
\begin{array}{c|c|c|c|c|r}
  A& B& C& D& E& \mathrm{Prob} \\
\hline
0& 0& 0& 0& 0&  \varepsilon \\
0& 0& 0& 0& 1&  1/4-\varepsilon \\
0& 1& 0& 0& 1&  1/4-\varepsilon \\
0& 1& 1& 0& 0&  \varepsilon \\
1& 0& 0& 0& 1&  1/4-\varepsilon \\
1& 0& 0& 1& 0&  \varepsilon \\
1& 1& 0& 0& 0&  \varepsilon \\
1& 1& 0& 0& 1&  1/4-\varepsilon
\end{array}
$$
For this particular distribution on $(A,B,C,D,E)$,
all terms of the dual quantity are zeroes except for two.
$I(C\lon D|AE) = 0$ since given any value of $(A,E)$, either $C$ or $D$ is
deterministic. A similar argument shows $I(C\lon D|BE) = 0$.
Given each value of $E$, the tuple $(A.B)$ is uniformly distributed among all
possible values, thus $I(A\lon B|E) = 0$.
To check that $I(A\lon E|CD) = 0$, we see that given some value of $(C,D)$ either $(A,E)$ is
deterministic or the distributions of $A$ and $E$ are independent (when $(C,D) =
(0,0)$). The case of $I(B\lon E|CD)$ is similar.
For the positive terms, we rely on formal computations which give:
\begin{align*}
  I(C\lon D|E) &= \Theta(\varepsilon),\\
  I(A\lon B|CD) &= \Theta(\varepsilon^2).
\end{align*}
Therefore the formal dual inequality cannot hold for small values of $\varepsilon$.
\end{IEEEproof}

We answer an open question of Matùš \cite{matus-asc}.
\begin{corollary}
  The dual of an almost representable matroid is not necessarily almost representable.
\end{corollary}
\begin{IEEEproof}
According to the main theorem, the entropy region is not closed under duality.
It implies there exist an entropic polymatroid $v$ whose dual is not almost entropic.
By \cite[Theorem~5]{matus-twocon}, there exist a sequence of entropic matroids 
that can asymptotically factor into a multiple of $v$ (by grouping elements).
Thus one of these entropic matroids must have a non-representable dual.

Notice that this proof is not constructive, however it can be made so by using
the explicit entropic polymatroid from the main theorem. Let us call it $v$, 
its dual polymatroid $v^\perp$ does not satisfy the MMRV inequality.
Thus we can construct an
entropic matroid whose dual is not entropic in the following way.  Approximate
$v$ by a (close enough) rational entropic vectors; use free expansion to
expand an integer multiple of $v$ into an entropic matroid $m$.  In
this way, $m$ is entropic but $m^\perp$ is not almost entropic: it fails the MMRV inequality. 
\end{IEEEproof}

The previous proof provides a construction for an entropic matroid whose dual
is not almost entropic, however no minimality claim is made. The smallest entropic matroid 
whose dual is not almost entropic is not known.

\section{Discussion}

This new geometrical property of the entropic region depicts a bigger geometric
picture. The entropic region is a cone that is not stable by duality and that is
stuck between two cones: the inner bound $\Hrank$ and the outer bound $\Hpoly$
that are both stable by duality. 
The case of duality of the entropy region for $n=4$ is still open. 

In future work, we investigate the case of information inequalities duality in
other settings, especially in quantum information inequalities \cite{polyquantoids}.
This notion of duality seems very general, in fact it applies to any concept
expressible via entropy. It can reveal dualities between
information theoretic problem, for instance the secret-sharing problem is
self-dual in the following sense. Any secret-sharing instance expressed with
entropy maps to another secret-sharing instance under the dual operator. 
In fact the dual instance is the secret-sharing problem for the dual access
structure.
In general, we expect a class of problems to be the dual of a different class of
problems.

\bibliography{bib}

\end{document}